\documentclass[fleqn,10pt]{wlscirep}
\usepackage{cite}

\newcommand{\etal}{\textit{et al. \ }}

\title{The Double Jones Birefringence in Magneto-electric Medium}

\author[1]{Waqas Mahmood}
\author[2,*]{Qing Zhao}

\affil[1,2]{School of Physics, Beijing Institute of Technology, $5$ South Zhongguancun Street, Haidian District, Beijing, $\mathit{100081}$, China}

\affil[*]{qzhaoyuping@bit.edu.cn}

\begin{abstract}
In this paper, the Maxwell's equations for a tensorial magneto-electric (ME) medium are solved, which is an extension to the work on the uniaxial anisotropic nonmagnetic medium. The coefficients of the dielectric permittivity, magnetic permeability, and of the magneto-electric effect are considered as tensors. The polarization is shown lying in the plane of two perpendicular independent vectors, and the relationship for the transverse polarization is given. The propagation of an electromagnetic wave through a ME medium gives rise to double Jones birefringence. Besides, the condition for an independent phenomenon of D'yakonov surface wave in a magneto-isotropic but with magneto-electric medium is given, which is measurable experimentally when the incident angle is $\frac{\pi}{4}$. Lastly, it is shown that the parameter for the magneto-electric effect plays a role in the damping of the wave.
\end{abstract}
\begin{document}

\flushbottom
\maketitle
%
%
\thispagestyle{empty}


\section*{Introduction}
\label{sec:0}
In $1888$, R\"{o}ntgen observed a connection between the electric and magnetic field by his observation, that a moving dielectric gets magnetized when it is placed in an electric field \cite{Rontgen}. His observation was followed by an entirely opposite phenomenon of the polarization of a moving dielectric in the presence of the magnetic field almost two decades later \cite{Wilson}. After these couple of findings, the inducement of polarization with magnetic field, and the inducement of magnetization with electric field became famous. The fact, that symmetry operations could be responsible for the coupling of both of these fields was raised by Curie for non-moving crystals \cite{Curie}. Though Curie realized, that his proposed intrinsic coupling based on symmetry operations between the fields is possible in non-moving crystals, but there was not enough explanation for that. Debye coined the magneto-electric (ME) effect \cite{Debye}, and years later Landau and Lifshitz proposed that the ME behavior is possible in time-asymmetric media \cite{LifshitzLandau}. This time-reversal symmetry was violated in antiferromagnetic $Cr_{2}O_{3}$ \cite{Dzyaloshinskii} and it was verified experimentally \cite{Astrov7, Astrov8, Rado9, Folen10}. In all observations, the electric field induced magnetization and the magnetic field induced polarization both are linear in the applied fields \cite{LifshitzLandau}. \\

The observations, that the electrostatic fields carry a link with optical effects have been discussed earlier. The most prominent of these optical effects is the linear birefringence that has been discussed by many authors \cite{Kerr1, Majorana11902, Majorana21902, CottonMouton1, CottonMouton2, CottonMouton3, CottonMouton4}. There has been lots of discussion on the media, that possibly show birefringence. From a popular calculus based formulism proposed by Jones to study the optical effects \cite{Jones1948}, it is clear that the uniaxial medium has the property of showing different fundamental optical effects such as isotropic refraction and absorption, linear birefringence and dichroism, and circular birefringence and dichroism. Another phenomenon initially predicted by Jones as the Jones effect was later observed experimentally by Roth et al. \cite{Roth}. \\

Now the idea has been extended to BiFeO$_{3}$ materials \cite{Yu} in which $D_{i}=\varepsilon_{ij}E_{j}+\alpha_{ij}H_{j}$ \cite{LifshitzLandau} (repeating indices mean summation). For antisymmetric $\mathbf{\alpha_{ij}}$ i.e., $\alpha_{ij}\thicksim\epsilon_{ijk}\upsilon_{k}$, it is equivalent to a moving medium with velocity $\upsilon_{i}$. The propagation of light in a moving medium has been extensively discussed in prior published articles \cite{Leonhardt, Slowlight}. Along a different line, very interesting developments have been made by setting $\alpha_{ij}\sim\theta(\vec{r}, t) \delta_{ij}$ ($\theta$ may depend on time or constant only). It leads to axion electrodynamics when $\theta$ is regarded as a dynamical variable \cite{Wilczek, Qi}, and gives rise to topological surface state related to the Lagrangian $\vec{E}\cdot\vec{B}$ with $\theta$ being constant. Besides, the relativistic nature of the magneto-electric modulus of Cr$_{2}$O$_{3}$ has been discussed by Heyl \etal and the four dimensional relativistic invariant pseudoscalar has been calculated \cite{FWHehl, ShimadaY}.\\

On the other hand, the propagation of electromagnetic (EM) wave in an anisotropic media has been widely investigated. Under the eikonal approximation $(\vec{E}=\vec{E}_{0} \ e^{\it{i}\psi}$, $\vec{k}=\nabla\psi$, and $\omega=-\frac{\partial\psi}{\partial t})$ \cite{LifshitzLandau}, many references are essential extension of the Fresnel's picture.  The article by Ignatovich \etal \cite{FVIgnatovich} and the references there in, reviewed and proposed the analytical description of EM waves in nonmagnetic anisotropic media by setting $D_{i}=\varepsilon_{ij}E_{j}$, where $\varepsilon_{ij}=\varepsilon_{1}\delta_{ij}+\varepsilon^{\prime}\hat{a}_{i}\hat{a}_{j}$. Here $a_{i}$ and $a_{j}$ are orthogonal unit vectors describing the anisotropic axes. $\varepsilon_{1}$ and $\varepsilon^{\prime}$ are the dielectric permittivity of isotropic and anisotropic media, and both are constants. This approach to describe the permittivity tensor using the addition of an axes is new, and it is used to study the optics of uniaxial anisotropic dielectric medium. The dispersion relation obtained from the Maxwell's equations for a nonmagnetic anisotropic medium has been studied in many aspects beyond the Fresnel's picture, and surface wave is proposed for certain angles \cite{FVIgnatovich}.\\

In this paper, we extend this approach to a uniaxial anisotropic magneto-electric (ME) medium. As it is prior mentioned, that the magneto-electric effect exists in linear relationship between the electric and magnetic fields in matter, therefore we introduce the same notation already given in Ref. \cite{FVIgnatovich} to describe our uniaxial anisotropic magneto-electric (ME) medium. The tensors $\mathbf{\epsilon_{ij}}$ and $\mathbf{\mu_{ij}}$ describe the dielectric permittivity and the magnetic permeability respectively. Obviously the tensor $\mathbf{\alpha_{ij}}$ in ME effect plays the role of anisotropic axes in the language of anisotropic media. Extending the idea given in Refs. \cite{Wilczek, Qi} with $\theta$ taken as a constant and $\mathbf{\alpha_{ij}}$ be considered as a symmetric constant tensor, we conclude that symmetric $\mathbf{\alpha_{ij}}$ under certain conditions when light is incident onto the ME surface gives rise to D'yakonov surface wave. We study how the ME effect terms appear in the final matrix, and what role they are playing in the underlying effect. We also propose the observation of surface wave under ME effect for certain angles and under special conditions.\\

The paper is organized as follows. In Section $2$, we discuss the calculations of the Maxwell's equations in a ME media, with constant tensors $\mathbf{\varepsilon_{ij}}$, $\mathbf{\mu_{ij}}$ and $\mathbf{\alpha_{ij}}$. We are interested in to find the polarization ($\vec{E}$) with respect to the equation $\nabla\cdot\vec{D}=0$. In Section $3$, some important cases are discussed with the solutions, the numerical plots of these results are given, and the expressions for permittivity and permeability matrix are compared with that given by Hehl \etal\cite{FWHehl}.

\section*{Magneto-electric (ME) effect in a magnetic uniaxial anisotropic medium}
\label{sec:A}
To transform the Maxwell's equations under a ME medium, we consider linear ME effect, and restrict our work to simpler terms by ignoring the higher order terms. Doing this, the typical relations for the ME effect take the form

\begin{equation} \label{MEeffectDi}
D_{i}= \epsilon_{ij} E_{j} + \alpha_{ij} H_{j} \ ,
\end{equation}

and

\begin{equation} \label{MEeffectBi}
B_{i}= \alpha_{ji} E_{j} + \mu_{ij} H_{j} \ .
\end{equation}

In the equations mentioned above, $\mathbf{\epsilon_{ij}}$ is the anisotropic dielectric permittivity and $\mathbf{\mu_{ij}}$ is the anisotropic magnetic permeability. The tensor $\mathbf{\alpha_{ij}}$ is the magneto-electric tensor and it is odd under time reversal \cite{LifshitzLandau}. The repeated indices mean summation. The simplest case is $\alpha_{ij} = \alpha(t) \delta_{ij}$ that has been studied extensively. In topological insulators, $\mathbf{\alpha}$ can be a constant rather than a dynamic field. Therefore, it is a natural extension that the tensor $\mathbf{\alpha_{ij}}$ is taken as a symmetric one.\\

The propagation of electromagnetic waves through any medium is described by the Maxwell's equations, and the behavior of the waves at the interface of two media is governed by the boundary conditions, imposed by these Maxwell's equations. Hence, the four Maxwell's equations for the case of no charges and current density can be written as

\begin{equation} \label{Maxwelleq1}
\nabla \cdot \vec{D}(\vec{r},t)=0 \ ,
\end{equation}

\begin{equation} \label{Maxwelleq2}
\nabla \cdot \vec{B}(\vec{r},t)=0 \ ,
\end{equation}

\begin{equation} \label{Maxwelleq3}
\nabla \times \vec{E}(\vec{r},t)= - \frac{\partial \vec{B} (\vec{r},t)}{\partial t} \ ,
\end{equation}

and

\begin{equation} \label{Maxwelleq4}
\nabla \times \vec{H}(\vec{r},t)= \frac{\partial \vec{D} (\vec{r},t)}{\partial t} \ .
\end{equation}

As it is mentioned earlier, that F. V. Ignatovich \etal have reviewed the case of a nonmagnetic anisotropic medium by introducing an additional axes to the dielectric permittivity, we here use the same technique of adding an additional axes to the dielectric permittivity and so on, to solve the case of a magnetic ME medium. This significant method of adding vectors make important the role of vectors.\\

Now consider a monochromatic (single frequency) wave of the form

\begin{equation} \label{eq:7}
\vec{E} (\vec{r},t)= \vec{E} (\vec{r}) \exp(-{\it{i} \omega t}) \ ,
\end{equation}

and further use an assumption, that no source ($\rho = 0$) and the current density ($\vec{j} = 0$) exists. Substituting Eq. (\ref{MEeffectDi}) and (\ref{MEeffectBi}) into the Maxwell's Eq. (\ref{Maxwelleq1}) - (\ref{Maxwelleq4}), we obtain

\begin{equation} \label{eq:8}
\nabla \times \Big[\mu^{-1} \cdot (\nabla \times \vec{E})\Big] - {\it{i}} \omega \Big\{\nabla \times \Big[(\mu^{-1} \alpha^{T}) \cdot \vec{E} \Big] - (\alpha \mu^{-1}) \cdot (\nabla \times \vec{E})\Big\} = \omega^2 \Big[\epsilon - (\alpha \mu^{-1} \alpha^{T})\Big] \cdot \vec{E} \ ,
\end{equation}

and

\begin{equation} \label{eq:9}
\vec{D} = \Big[\epsilon - (\alpha \mu^{-1} \alpha^T)\Big]\cdot \vec{E}+\frac{1}{\it{i} \omega}(\alpha \mu^{-1}) \cdot (\nabla \times \vec{E}) \ ,
\end{equation}

where $\mathbf{\epsilon}$, $\mathbf{\mu^{-1}}$ and $\mathbf{\alpha}$ represent the tensors with matrix elements $\mathbf{\epsilon_{ij}}$, $\mathbf{(\mu^{-1})_{ij}}$ and $\mathbf{\alpha_{ij}}$. Since, we are dealing with an anisotropic medium, and for the characterization of anisotropic medium we consider $\epsilon$, $\mu$ and $\alpha$ as symmetric tensors. Now setting

\begin{equation} \label{eq:10}
\beta = \alpha \mu^{-1} \ ,
\end{equation}

and

\begin{equation} \label{eq:11}
\tilde{\varepsilon}_{ij} = \epsilon_{ij}-(\beta \alpha^T)_{ij} \ ,
\end{equation}

in Eq. (\ref{eq:8}) and (\ref{eq:9}), we obtain

\begin{equation} \label{eq:12}
\nabla \times \Big[\mu^{-1} \cdot (\nabla \times \vec{E})\Big] - {\it{i}} \omega \Big\{\nabla \times (\beta \cdot \vec{E}) - \beta \cdot (\nabla \times \vec{E})\Big \} = \omega^{2} \tilde{\varepsilon} \cdot \vec{E} \ ,
\end{equation}

and

\begin{equation} \label{eq:13}
\vec{D} = \tilde{\varepsilon} \cdot \vec{E}+\frac{1}{\it{i} \omega}\beta \cdot (\nabla \times \vec{E}) \ .
\end{equation}

Next, we extend the anisotropic dielectric medium to an anisotropic magneto-electric medium. For such a medium, the isotropic dielectric permittivity is denoted by $\epsilon_{1}$, the strength of the anisotropy is represented as $\epsilon'$, the inverse of isotropic magnetic permeability is $\tau_{1}$, the inverse of anisotropic magnetic permeability is taken as $\tau'$, $\beta_{1}$ is the isotropic ME coefficient and $\beta'$ is the anisotropic ME coupling coefficient. Taking the role of additional vectors into account, and writing all the constant coefficients $\mathbf{\epsilon_{1}}, \mathbf{\epsilon'}, \mathbf{\tau_{1}}, \mathbf{\tau'}, \mathbf{\beta_{1}}$ and $\mathbf{\beta'}$ in the tensors, the anisotropic medium takes the form

\begin{equation} \label{eq:14}
\tilde{\varepsilon}_{ij} = \epsilon_{1} \delta_{ij} + \epsilon' \hat{a_{i}}\hat{a_{j}} \ ,
\end{equation}

\begin{equation} \label{eq:15}
(\mu^{-1})_{ij} = \tau_{1} \delta_{ij} + \tau' \hat{b_{i}}\hat{b_{j}} \ ,
\end{equation}

and

\begin{equation} \label{eq:16}
\beta_{ij} = \beta_{1} \delta_{ij} + \beta' \hat{d_{i}}\hat{d_{j}} \ ,
\end{equation}

where the relationships for $\beta_{1}$, $\beta'$ and the matrix form of $\alpha_{ij}$ usually used in experiments shown in Ref. \cite{FWHehl} will be given later. The orthogonal unit vectors $\hat{a}$, $\hat{b}$ and $\hat{d}$ introduced as additional axes in Eq. (\ref{eq:14})--(\ref{eq:16}), are given by

\begin{equation} \label{eq:17}
\hat{a} = (\sin\theta \cos\phi, \hspace{0.05cm}\sin\theta \sin\phi, \hspace{0.05cm} \cos\theta ) \ ,
\end{equation}

\begin{equation} \label{eq:18}
\hat{b} = (\sin\varphi \cos\psi,\hspace{0.05cm} \sin\varphi \sin\psi,\hspace{0.05cm} \cos\varphi ) \ ,
\end{equation}

and

\begin{equation} \label{eq:19}
\hat{d} = (\sin\beta \cos\delta, \hspace{0.05cm}\sin\beta \sin\delta, \hspace{0.05cm} \cos\beta ) \ .
\end{equation}

It is important to note that the angle $\beta$ in Eq. (\ref{eq:19}) is different from the parameters $\beta_{1}$ and $\beta'$ in Eq. (\ref{eq:16}). As a result of an added axes, we expect the off diagonal terms of the final matrix to be non-zero, which in the case of nonmagnetic medium were equal to zero. The coefficient $\beta'$ appearing with the additional axes is assumed to play a crucial role. For convenience, we choose $\hat{\kappa} = \frac{\vec{k}}{\it k}$ along $z-axes$ to completely describe our system, and introduce a new set of vectors in the form 

\begin{equation} \label{eq:20}
\vec e_{1} = \hat{a} \times \hat{\kappa} \ ,
\end{equation}

\begin{equation} \label{eq:21}
\vec e = \hat{b} \times \hat{\kappa} \ ,
\end{equation}

and

\begin{equation} \label{eq:22}
\vec g = \hat{d} \times \hat{\kappa} \ ,
\end{equation}

where $e$ results in replacing $\theta$ and $\phi$ by $\varphi$ and $\psi$. \\

For the plane wave, we can substitute $\nabla \rightarrow {\it{i}} \vec{k}$. Using this substitution in Eq. (\ref{Maxwelleq1}), and further substituting Eq. (\ref{eq:13}) into it, we arrive at

\begin{equation} \label{eq:23}
\epsilon_{1}(\kappa \cdot E) + \epsilon' (\kappa \cdot a) (a \cdot E) + \xi \beta' (\kappa \cdot d) (g \cdot E) = 0 \ ,
\end{equation}

where $\frac{\omega}{c} = k_{0}$, $\vec g = (\hat d \times \hat\kappa)$ and $\xi = \frac{\it k}{k_{0}} = \frac{c\it k}{\omega}$.\\

Now substituting $\nabla \rightarrow {\it{i}} \vec{k}$ into Eq. (\ref{eq:12}), and further solving it after using Eq. (\ref{eq:21}) into it, we obtain

\begin{equation} \label{eq:24}
(\tau_{1}\xi^{2} - \epsilon_{1})E - \tau_{1}\xi^{2} (\kappa \cdot E) \kappa + \tau' \xi^{2} (e \cdot E)e - \epsilon' (a \cdot E) a - \beta' \xi \Big[g(\hat d \cdot E) + \hat d (g \cdot E)\Big] = 0 \ .
\end{equation}

In the new basis,

\begin{equation} \label{eq:25}
\vec{E} = A \hat{a} + B \hat{\kappa} + C \vec e_{1} \ ,
\end{equation}

where $\hat{a}$, $\hat{\kappa}$ and $\vec e_{1}$ form a right handed system. The coordinates A, B, and C are not independent. In order to find the value of the coordinate B,  we substitute Eq. (\ref{eq:25}) into Eq. (\ref{eq:23}) ($\nabla \cdot \vec{D} = 0$), and obtain

\begin{equation} \label{eq:26}
\begin{split}
B = & \frac{-1}{(1+ \eta \cos^2 \theta)}\Bigg[ A \bigg \{(1+\eta) \cos\theta - \rho \xi \sin\theta \cos\beta \sin\beta \sin(\phi-\delta) \bigg \} + C \bigg \{\rho \xi \sin\theta \cos\beta \sin\beta \\ & \cos(\phi-\delta) \bigg \}\Bigg] \ ,
\end{split}
\end{equation}

where $\rho = \frac{\beta'}{\epsilon_{1}}$, $\eta = \frac{\epsilon'}{\epsilon_{1}}$ and $\xi = \frac{k}{k_{0}}$. \\

Substituting $B$ into Eq. (\ref{eq:25}), we obtain the polarization vector in the form

\begin{equation} \label{eq:27}
\vec{E}= A \vec{e_{2}} + C \vec{e_{1}} \ ,
\end{equation}

where

\begin{equation} \label{eq:28}
\vec e_{2} = \hat a - \frac{\hat\kappa (1 + \eta) \cos\theta}{(1+ \eta \cos^2 \theta)}\Big(1 + \xi \chi\Big) \ ,
\end{equation}

and

\begin{equation} \label{eq:29}
 \chi= - \frac{\rho}{(1 + \eta)} \tan\theta \cos\beta \sin\beta \sin(\phi - \delta) \Big(1 - \frac{C}{A} \cot(\phi - \delta)\Big) \ .
\end{equation}

\begin{figure}[H]
\begin{center}
\includegraphics[scale=0.1]{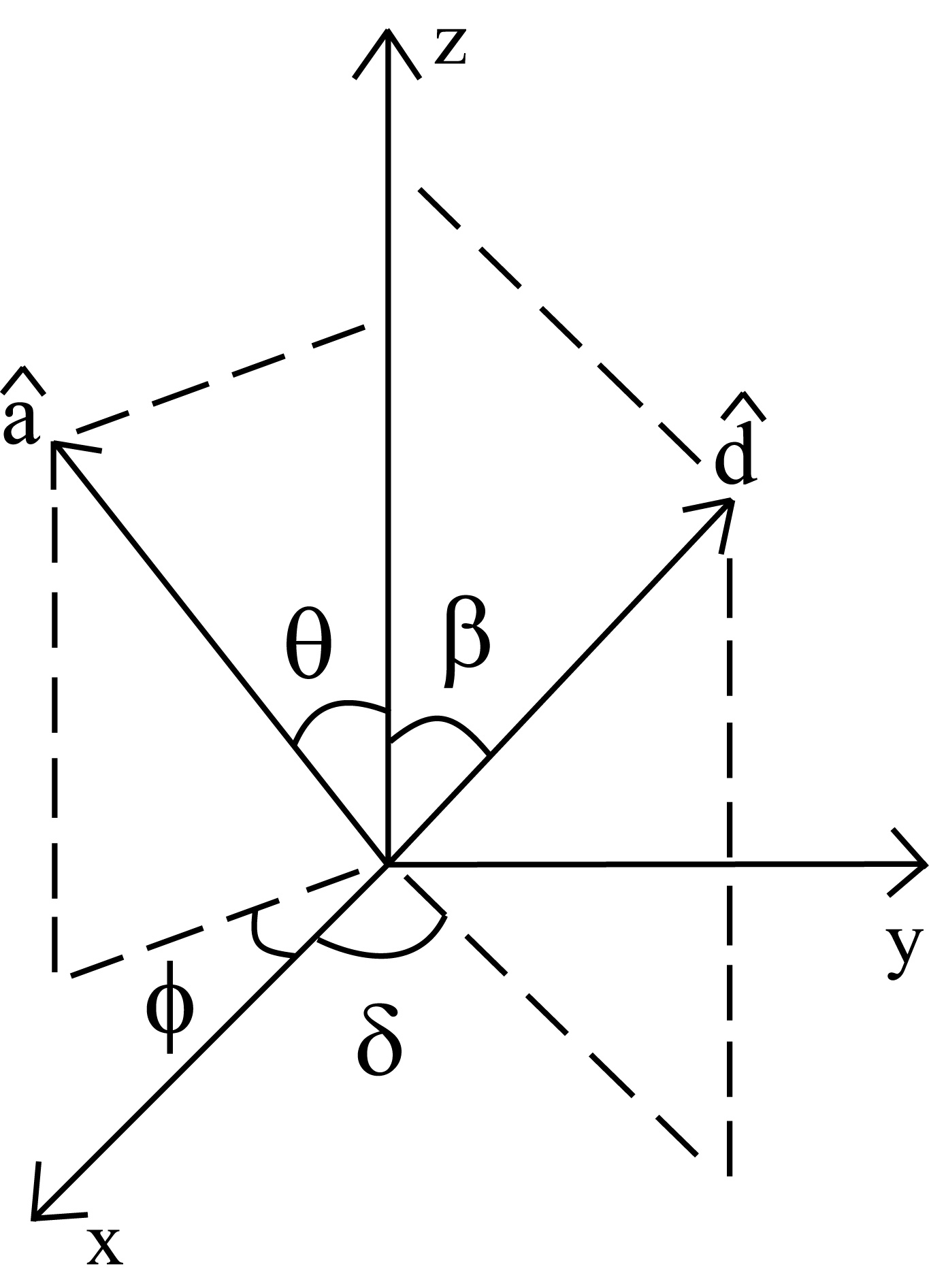}
\caption{The unit vectors $\hat a$  and $\hat d$ for the case when $\phi - \delta = \frac{\pi}{2}$.}
\label{orthogonal-plane}
\end{center}
\end{figure}

It is evident from Eq. (\ref{eq:27}), that the polarization vector ($\vec E$) lies in the plane of two independent orthogonal vectors $\vec e_{1}$ and $\vec e_{2}$, where $\vec e_{2}$ is given by Eq. (\ref{eq:28}). The transverse polarization can be found by replacing the vector $\hat{a}$ given in Eq. (\ref{eq:25}) by $\hat{a_{t}} = \hat{a}-\hat\kappa(\hat\kappa \cdot \hat{a})$. In order to deal with the linear theory, we consider an interesting case when $ \phi - \delta= \frac{\pi}{2}$ i.e., the plane of $\hat{a}$ and $\hat{\kappa}$, and the plane of $\hat{d}$ and $\hat{\kappa}$ are perpendicular to each other as shown in Figure (\ref{orthogonal-plane}) .

Hence Eq. (\ref{eq:29}) implies

\begin{equation} \label{eq:30}
\chi = - \frac{\rho}{(1 + \eta)} \tan\theta \cos\beta \sin\beta \ .
\end{equation}

Substituting $\chi$ back into Eq. (\ref{eq:28}), we have

\begin{equation} \label{eq:31}
\vec e_{2} = \hat a - \hat\kappa \frac{\epsilon_{2}(\theta)}{\epsilon_{1}} \cos\theta \Big(1 - \frac{\rho \xi}{(1 + \eta)} \tan\theta \cos\beta \sin\beta\Big) \ ,
\end{equation}

where

\begin{equation} \label{eq:32}
\epsilon_{2}(\theta) = \frac{\epsilon_{1} (1+ \eta)}{(1+ \eta \cos^2 \theta)} \ .
\end{equation}

In terms of

\begin{equation} \label{eq:33}
\lambda_{1} = \frac{\epsilon_{2}(\theta)}{\epsilon_{1}} \cos\theta \Big(1 - \frac{\rho \xi}{(1+\eta)} \tan\theta \cos\beta \sin\beta\Big) \ ,
\end{equation}

we have

\begin{equation} \label{eq:34}
\vec e_{2} = \hat a - \lambda_{1} \hat\kappa \ ,
\end{equation}

where it is important to note that $\vec{e_{1}} \cdot \vec{e_{2}} = 0$ \ . \\

The meaning of Eq. (\ref{eq:34}) is obvious. Because of Eq. (\ref{Maxwelleq1}), the transverse condition has been satisfied. Therefore, $\vec{E}$ has two independent polarizations.\\

Multiplying  Eq. (\ref{eq:24}) by $\vec e_{1}$ yields

\begin{equation} \label{eq:35}
C \hspace{0.1cm} \Big(\tau_{1}\xi^{2} - \epsilon_{1}\Big) + A \hspace{0.1cm} \Big(\beta' \xi \sin^2\beta\Big) = 0 \ .
\end{equation}

Now multiplying  Eq. (\ref{eq:24}) by $\vec e_{2}$, using Eq. (\ref{eq:33}) into it, and performing a little lengthy calculations we obtain

\begin{equation} \label{eq:36}
C \hspace{0.1cm} \Big(\beta' \xi \sin^2\beta\Big) + A \hspace{0.1cm} \Big(\Big\{\tau_{1} + F + \tau' \sin^2\varphi\Big\}\xi^2 - \epsilon_{2}(\theta) - G \xi\Big) = 0 \ ,
\end{equation}

where

\begin{equation} \label{eq:37}
F = \Big(\frac{\beta'}{\epsilon_{1}}\Big)^2 \cos^2\beta \sin^2\beta \frac{\epsilon_{2}(\theta)}{(1+\eta)} \ ,
\end{equation}

and

\begin{equation} \label{eq: 38}
G = \eta \rho \sin2\beta \cos\theta \sin\theta \frac{\epsilon_{2}(\theta)}{(1 + \eta)} \ .
\end{equation}

The Eq. (\ref{eq:35}) and (\ref{eq:36}) can be written in the matrix form as

\begin{equation}\label{eq:matrix1}
\Bigg[ \begin{array}{cc} \tau_{1}\xi^{2} - \epsilon_{1} & \beta' \xi \sin^2\beta \\
\beta' \xi \sin^2\beta & (\tau_{1} + F + \tau' \sin^2\varphi)\xi^2 - \epsilon_{2}(\theta) - G \xi \end{array} \Bigg] \Bigg[ \begin{array}{c} C \\
A \end{array} \Bigg] = 0 \ .
\end{equation}

Eq. (\ref{eq:matrix1}) is the general form of the propagation of transversal EM wave in a ME media for $\phi-\delta=\frac{\pi}{2}$. We shall discuss the meaning of Eq. (\ref{eq:matrix1}) in Section 3. It should be noted that for DC ($\omega\rightarrow0$), $\xi$ is very large. However, for AC, we consider low frequencies and the correction of linear terms of $\xi$. In our work, the light propagation is taken into account, however, the theory works for any frequency in principle.

\section*{Particular cases and proposed surface wave}
\label{sec:B}

Let's start with a trivial case, when there is no magneto-electric effect i.e., when $\beta' = \tau' = 0$. The matrix given in Eq. (\ref{eq:matrix1}) reduces to

\begin{equation}
\Bigg[ \begin{array}{cc} \tau_{1} \xi^{2} - \epsilon_{1} & 0 \\
0 & \tau_{1} \xi^2 - \epsilon_{2}(\theta) \end{array} \Bigg] \Bigg[ \begin{array}{c} C \\
A \end{array} \Bigg] = 0 \ ,
\end{equation}

which can be written in the equations form for $\tau_{1} = 1$ as

\begin{equation} \label{eq:39}
\xi^{2} = \epsilon_{1} \ ,
\end{equation}

and

\begin{equation} \label{eq:40}
\xi^{2} = \epsilon_{2}(\theta) \ .
\end{equation}

The Eq. (\ref{eq:39}) and (\ref{eq:40}) are nothing but the relationships of the wave vector for an anisotropic dielectric medium already discussed in \cite{FVIgnatovich}, when there exists only one anisotropic axes $\hat a$. When tensorial ME effect is considered, the off diagonal terms are non zero, however, the term $\beta' \xi \sin^2\beta$ appears in the off diagonal term's place, which is constrained by the Maxwell's Eq. (\ref{Maxwelleq3}) and (\ref{Maxwelleq4}).\\

Another interesting and special case is, when $\eta = 0$ i.e., the medium is with magnetic structure and magneto-electric effect only. Rewriting Eq. (\ref{eq:matrix1}) using this assumption, we have

\begin{equation} \label{matrix41}
\Bigg[ \begin{array}{cc} \tau_{1}\xi^{2} - \epsilon_{1} & \beta' \xi \sin^2\beta \\
\beta' \xi \sin^2\beta & \nu \xi^2 - \epsilon_{1} \end{array} \Bigg] \Bigg[ \begin{array}{c} C \\
A \end{array} \Bigg] = 0 \ ,
\end{equation}

where

\begin{equation} \label{eq:G-eta-assumption}
G = 0 \ ,
\end{equation}

\begin{equation} \label{eq:epsilon2theta-eta-assumption}
\epsilon_{2}(\theta) = \epsilon_{1} \ ,
\end{equation}

\begin{equation} \label{eq:nu-eta-assumption}
\nu = (\tau_{1} + F + \tau' \sin^2\varphi) \ ,
\end{equation}

and

\begin{equation} \label{eq:F-eta-assumption}
F =  \Big(\frac{\beta'}{\epsilon_{1}}\Big)^2 \cos^2\beta \sin^2\beta \epsilon_{1} \ .
\end{equation}

To find the dispersion relation satisfied by the matrix given in Eq. (\ref{matrix41}), we set the determinant of the matrix equal to zero, and arrive at

\begin{equation} \label{eq:dispersion-eta-assumption}
\tau_{1} \nu \xi^4 - \Big(\tau_{1} \epsilon_{1} + \nu \epsilon_{1} + \beta'^2 \sin^4\beta\Big) \xi^2 + \epsilon_{1}^2 = 0 \ .
\end{equation}

Recalling

\begin{equation} \label{eq:xi-etta-assumption}
\xi = \frac{k}{k_{0}} \ ,
\end{equation}

and denoting

\begin{equation} \label{eq:ksquare-etta-assumption}
k^2 = k_{\parallel}^2 + k_{\perp}^2 \ ,
\end{equation}

where $k_{\parallel}$ denotes the parallel component, and $k_{\perp}$ denotes the perpendicular component to the boundary $z=0$ of the media. If light is incident on the boundary, $k_{\parallel}$ remains unchanged for $z>0$ and $z<0$ because of the continuity. The Eq. (\ref{eq:dispersion-eta-assumption}) recasts to

\begin{equation} \label{eq:fourth-order-etta-assumption}
\tau_{1} \nu k_{\perp}^4 + \Big(2 k_{\parallel}^2 \tau_{1} \nu - k_{0}^2 D \Big) k_{\perp}^2 + k_{0}^4 \epsilon_{1}^2 + \tau_{1} \nu k_{\parallel}^4 - k_{\parallel}^2 k_{0}^2 D = 0 \ ,
\end{equation}

where

\begin{equation} \label{eq:D-etta-assumption}
D = \tau_{1} \epsilon_{1} + \nu \epsilon_{1} + \beta'^2 \sin^4\beta \ .
\end{equation}

Solving Eq. (\ref{eq:fourth-order-etta-assumption}) as a quadratic equation in $k_{\perp}^2$, we obtain the solution

\begin{equation} \label{eq:solution-etta-zero}
k_{\perp}^2 = \frac{k_{0}^2}{2 \tau_{1} \nu} \Big(D \pm \sqrt{D^2 - 4 \tau_{1} \nu \epsilon_{1}^2}\Big) - k_{\parallel}^2 \ .
\end{equation}

The component $k_{\parallel}^{2}$ is invariant and the squared term $D^2$ in  Eq. (\ref{eq:solution-etta-zero}) is always larger than the second term in the square root. The dispersion relation depends on the axes of magneto-electric effect and for negative $k_{\parallel}^{2}$, only surface wave is survived.

\subsection*{Surface wave in uniaxial anisotropic ME medium}
\label{sec:D}

\begin{figure}[H]
\begin{center}
\includegraphics[scale=0.38]{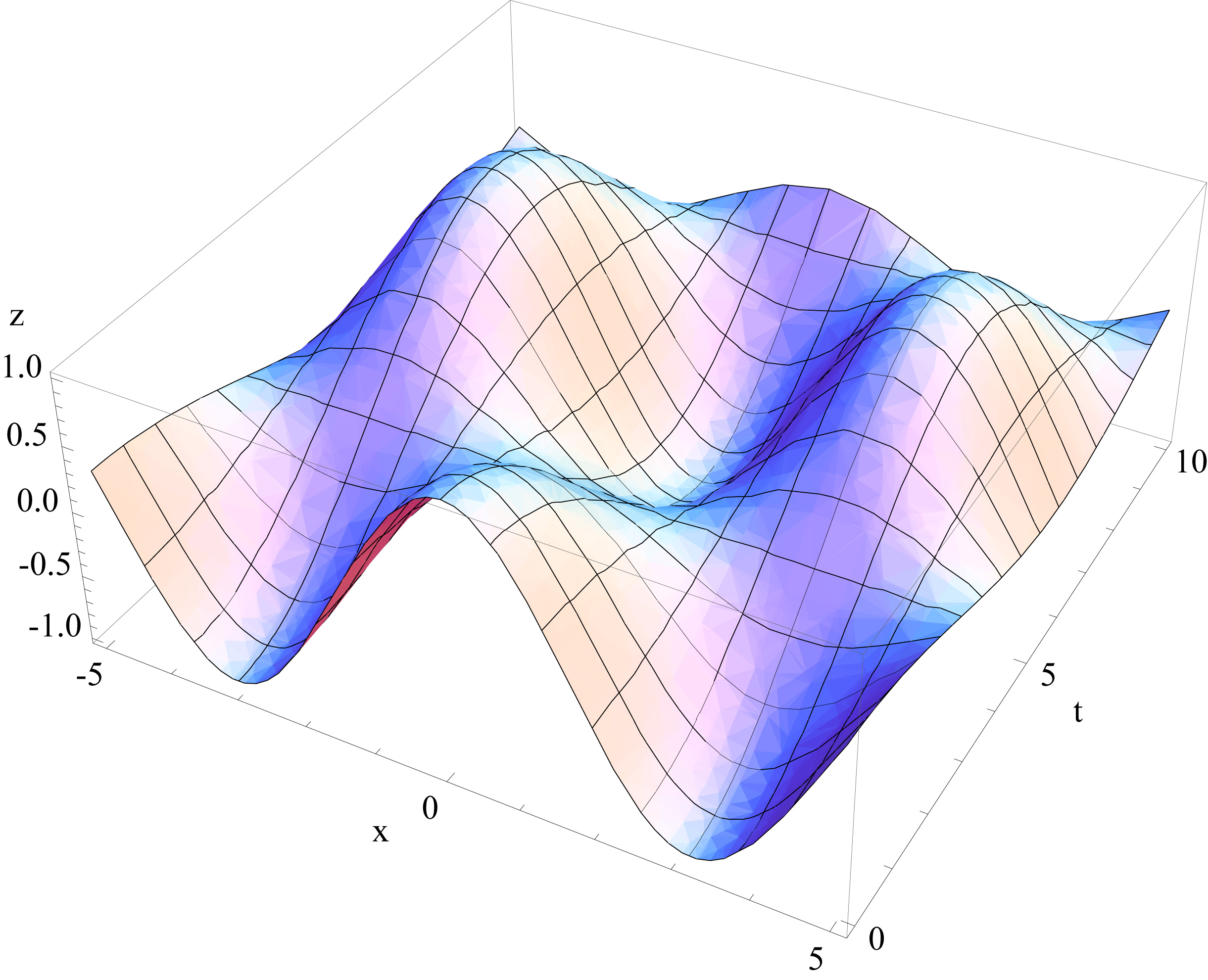}
\caption{Numerical plot of an electromagnetic wave with time harmonic form and $exp(\textit{-i} \omega t)$ dependence.}
\label{emw-plot}
\end{center}
\end{figure}

It is mentioned earlier, that surface wave exists in a uniaxial anisotropic dielectric medium under certain incident angles \cite{FVIgnatovich}. Consider now the case of a uniaxial anisotropic magneto-electric medium. Let's consider a region of two halves separated by the plane $z = 0$. The region $z > 0$ is vacuum, and the region $z < 0$ is the magneto-electric medium. Due to the continuity of $\vec{E}$, the parallel component ($k_{\parallel}$) does not change. However, the only change occurs in the perpendicular component ($k_{\perp})$. Suppose a plain wave propagates towards a plane at $z = 0$ (Figure. \ref{emw-plot}).

For the general case, the roots of the fourth order algebraic equation looks complicated. Hence, we consider a special case with $\eta \neq 0$ and $\theta = \frac{\pi}{2}$. For this case, the anisotropic axes for the dielectric tensor mentioned in Eq. (\ref{eq:14}) is perpendicular to the propagation direction $\hat\kappa$, or only an incident wave propagating along the direction perpendicular to $\hat{a}$ is taken into account.\\

Substituting $\theta = \frac{\pi}{2}$ in the Eq. (\ref{eq:matrix1}), it is simplified to\\

\begin{equation} \label{matrix432}
\Bigg[ \begin{array}{cc} \tau_{1}\xi^{2} - \epsilon_{1} & \beta' \xi \sin^2\beta \\
\beta' \xi \sin^2\beta & \nu \xi^2 - \epsilon_{1} (1 + \eta) \end{array} \Bigg] \Bigg[ \begin{array}{c} C \\
A \end{array} \Bigg] = 0 \ ,
\end{equation}

where

\begin{equation}
\nu = \tau_{1} + F + \tau' \sin^{2}\varphi \ ,
\end{equation}

and

\begin{equation}
F = \bigg(\frac{\beta'}{\epsilon_{1}}\bigg)^2 \cos^{2}\beta \sin^{2}\beta \epsilon_{1} \ .
\end{equation}

\subsection*{Magneto-isotropic but with magneto-electric tensor}
\label{sec:E}

Let's consider another special case in which $\hat b$ and $\alpha$ both don't play any role. However, only $\hat d$ is survived. Suppose $\epsilon' = 0$ and $\tau' = 0$. The matrix in Eq. (\ref{eq:matrix1}) becomes

\begin{equation} \label{matrix23}
\Bigg[ \begin{array}{cc} \tau_{1}\xi^{2} - \epsilon_{1} & \beta' \xi \sin^2\beta \\
\beta' \xi \sin^2\beta & \nu \xi^2 - \epsilon_{1} \end{array} \Bigg] \Bigg[ \begin{array}{c} C \\
A \end{array} \Bigg] = 0 \ ,
\end{equation}

where

\begin{equation}
G = 0 \ ,
\end{equation}

\begin{equation}
\epsilon_{2}(\theta) = \epsilon_{1} \ ,
\end{equation}

\begin{equation}
\rho^2 = \Big(\frac{\beta'}{\epsilon_{1}}\Big)^2 \ ,
\end{equation}

\begin{equation}
\nu = (\tau_{1} + F) \ ,
\end{equation}

and

\begin{equation}
F = \rho^2 \cos^2\beta \sin^2\beta \epsilon_{1} \ .
\end{equation}

Taking the determinant of the matrix given in Eq. (\ref{matrix23}), and setting it equal to zero, we arrive at

\begin{equation} \label{eq:obtainedfromM4}
\tau_{1} \nu \xi^4 - \Big(\tau_{1}\epsilon_{1} + \nu \epsilon_{1} + \beta'^2 \sin^4\beta\Big) \xi^2 + \epsilon_{1}^2 = 0 \ .
\end{equation}

Using $\xi = \frac{k}{k_{0}}$, and further solving Eq. (\ref{eq:obtainedfromM4}), we arrive at

\begin{equation} \label{eq:tau1k4}
\tau_{1} k^4 + \rho^2 \Big(k^2 \cos^2\beta\Big) \Big(k^2 \sin^2\beta\Big) \epsilon_{1} - k_{0}^2 \Big(2 \epsilon_{1} k^2 + \frac{\beta'^2}{\tau_{1}} k^2 \sin^2\beta\Big) + k_{0}^4 \frac{\epsilon_{1}^2}{\tau_{1}} = 0 \ .
\end{equation}

Introducing the components form, we have

\begin{equation} \label{eq:k-cos-beta}
k \cos\beta = k_{\parallel} (\hat l \cdot \hat d) + k_{\perp} (\hat n \cdot \hat d) \ ,
\end{equation}

which on squarring both sides yields

\begin{equation} \label{eq: k2-cos2-beta}
k^2 \cos^2\beta = k_{\parallel}^2 ( \hat l \cdot \hat d)^2 + 2 k_{\parallel} k_{\perp} ( \hat l \cdot \hat d) ( \hat n \cdot \hat d) + k_{\perp}^2 (\hat n \cdot \hat d)^2 \ ,
\end{equation}

where $\hat l$ is the unit vector in the direction parallel to the surface and $\hat n$ is the unit vector in the direction perpendicular to the surface. \\

Similarly

\begin{equation} \label{eq:k2-sin2-beta}
k^2 \sin^2\beta = k_{\parallel}^2 \Big(1  - ( \hat l \cdot \hat d)^2 \Big) + k_{\perp}^2 \Big(1 - ( \hat n \cdot \hat d)^2\Big) - 2 k_{\parallel} k_{\perp} ( \hat l \cdot \hat d) ( \hat n \cdot \hat d) \ .
\end{equation}

Multiplying Eq. (\ref{eq: k2-cos2-beta}) and (\ref{eq:k2-sin2-beta}), we have

\begin{equation} \label{eq:product}
\begin{split}
&\Big(k^2 \cos^2\beta\Big) \Big(k^2 \sin^2\beta\Big) = k_{\perp}^4 \Big((\hat n \cdot \hat d)^2 - (\hat n \cdot \hat d)^4\Big) + k_{\perp}^3 \Big(2 (\hat l \cdot \hat d) (\hat n \cdot \hat d) k_{\parallel} - 4 (\hat l \cdot \hat d)(\hat n \cdot \hat d)^3 k_{\parallel} \Big) \\ & + k_{\perp}^2 \Big((\hat l \cdot \hat d)^2 k_{\parallel}^2 + (\hat n \cdot \hat d)^2 k_{\parallel}^2 - 6 (\hat l \cdot \hat d)^2 (\hat n \cdot \hat d)^2  k_{\parallel}^2 \Big) + k_{\perp} \Big(2 (\hat l \cdot \hat d) (\hat n \cdot \hat d) k_{\parallel}^3 - 4 (\hat l \cdot \hat d)^3 (\hat n \cdot \hat d) k_{\parallel}^3\Big) \\ & + \Big((\hat l \cdot \hat d)^2 - (\hat l \cdot \hat d)^4\Big) k_{\parallel}^4 \ .
\end{split}
\end{equation}

Recall

\begin{equation} \label{eq: ksquare}
k^2 = k_{\parallel}^2 + k_{\perp}^2 \ ,
\end{equation}

and likewise

\begin{equation} \label{eq:k4}
k^4 = k_{\parallel}^4 + 2 k_{\parallel}^2 k_{\perp}^2 + k_{\perp}^4 \ .
\end{equation}

Let $x = k_{\perp}$, and substitute Eq. (\ref{eq:product}), (\ref{eq: ksquare}) and (\ref{eq:k4}) into Eq. (\ref{eq:tau1k4}). We arrive at the following fourth order algebraic equation in $x$

\begin{equation} \label{fulleqn}
\begin{split}
\Big(&\tau_{1} + \cos^2\gamma \sin^2\gamma \hspace{0.1cm} \frac{\beta'^2}{\epsilon_{1}}\Big) x^4 + \Big(- \frac{\beta'^2}{\epsilon_{1}} k_{\parallel} \cos2\gamma \sin2\gamma\Big) x^3 + \Big(k_{\parallel}^2 \Big(2 \tau_{1} + \frac{\beta'^2}{\epsilon_{1}} (1 - 6 \cos^2\gamma \sin^2\gamma)\Big)\\ & - k_{0}^2 \Big(2 \epsilon_{1} + \frac{\beta'^2}{\tau_{1}} \sin^2\gamma\Big)\Big) x^2 + \Big(\frac{\beta'^2}{\epsilon_{1}} k_{\parallel}^3 \cos2\gamma \sin2\gamma + \frac{\beta'^2}{\tau_{1}} k_{0}^2 k_{\parallel} \sin2\gamma \Big) x + \Big(k_{\parallel}^4 \Big(\tau_{1} + \frac{\beta'^2}{\epsilon_{1}} \cos^2\gamma \\& \sin^2\gamma\Big) - k_{0}^2 k_{\parallel}^2 \Big(2 \epsilon_{1} + \frac{\beta'^2}{\tau_{1}} \cos^2\gamma\Big) + \frac{k_{0}^4 \epsilon_{1}^2}{\tau_{1}}\Big) = 0 \ ,
\end{split}
\end{equation}

where $\gamma$ is the angle between $\hat n$ and $\hat d$. The equation given above can be rewritten as

\begin{equation} \label{ax4}
a x^4 + b x^3 + c x^2 + d x + e = 0 \ ,
\end{equation}

where

\begin{equation}
a = \tau_{1} + \cos^2\gamma \sin^2\gamma \hspace{0.1cm} \frac{\beta'^2}{\epsilon_{1}} \ ,
\end{equation}

\begin{equation}
b = - \frac{\beta'^2}{\epsilon_{1}} k_{\parallel} \cos2\gamma \sin2\gamma \ ,
\end{equation}

\begin{equation}
c = k_{\parallel}^2 \Big(2 \tau_{1} + \frac{\beta'^2}{\epsilon_{1}} (1 - 6 \cos^2\gamma \sin^2\gamma)\Big) - k_{0}^2 \Big(2 \epsilon_{1} + \frac{\beta'^2}{\tau_{1}} \sin^2\gamma\Big) \ ,
\end{equation}

\begin{equation}
d = \frac{\beta'^2}{\epsilon_{1}} k_{\parallel}^3 \cos2\gamma \sin2\gamma + \frac{\beta'^2}{\tau_{1}} k_{0}^2 k_{\parallel} \sin2\gamma \ ,
\end{equation}

and

\begin{equation}
e = k_{\parallel}^4 \Big(\tau_{1} + \frac{\beta'^2}{\epsilon_{1}} \cos^2\gamma \sin^2\gamma\Big) - k_{0}^2 k_{\parallel}^2 \Big(2 \epsilon_{1} + \frac{\beta'^2}{\tau_{1}} \cos^2\gamma\Big) + \frac{k_{0}^4 \epsilon_{1}^2}{\tau_{1}} \ .
\end{equation}

\begin{figure}[H]
\begin{center}
\includegraphics[scale=0.08]{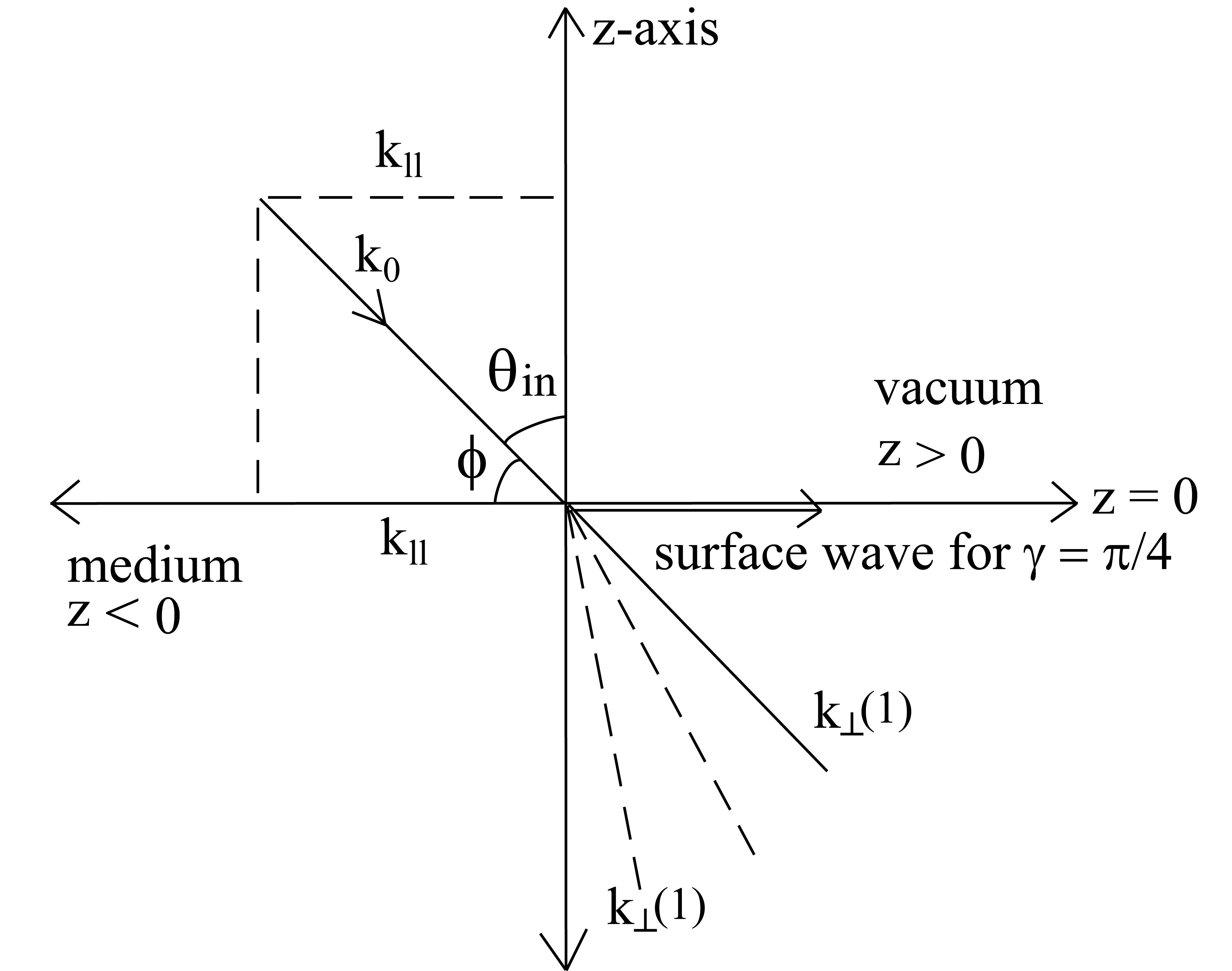}
\caption{The region z $>$ 0 is vacuum and the region z $<$ 0 is magneto-electric medium. $\theta_{in}$ is the incident angle for $k_{0}$. The parallel and perpendicular components are $k_{\parallel}$ and $k_{\perp}$ respectively. The range of $\theta_{in}$ is constrained by $\beta$ and for $\gamma = \frac{\pi}{4}$ surface wave along the plane z = 0 is shown.}
\label{SW}
\end{center}
\end{figure}

The general solution of the fourth order equation given in Eq. (\ref{ax4}) is very complicated. To show the point, we consider a simple example for $\gamma=\frac{\pi}{4}$ (Figure. \ref{SW}), that can be checked experimentally in principle \cite{Roth}.

Setting $\gamma = \frac{\pi}{4}$ and finding the solutions, we have

\begin{equation}\label{eq:S-1}
k_{\perp(1)} = \frac{1}{\sqrt{4 \epsilon_{1} + \beta'^2}} \Bigg[k_{\parallel} \beta' - \sqrt{\epsilon_{1} k_{0}^2 \Big(4 \epsilon_{1} + \beta'^2 - \beta' \sqrt{4 \epsilon_{1} + \beta'^2} \Big) - k_{\parallel}^2 (4\epsilon_{1})} \hspace{0.2cm}\Bigg] \ ,
\end{equation}

\begin{equation}\label{eq:S-2}
k_{\perp(2)} = \frac{1}{\sqrt{4 \epsilon_{1} + \beta'^2}} \Bigg[k_{\parallel} \beta' + \sqrt{\epsilon_{1} k_{0}^2 \Big(4 \epsilon_{1} + \beta'^2 - \beta' \sqrt{4 \epsilon_{1} + \beta'^2}\Big) - k_{\parallel}^2 (4\epsilon_{1})} \hspace{0.2cm}\Bigg] \ ,
\end{equation}

\begin{equation}\label{eq:S-3}
k_{\perp(3)} = \frac{1}{\sqrt{4 \epsilon_{1} + \beta'^2}} \Bigg[- k_{\parallel} \beta' - \sqrt{\epsilon_{1} k_{0}^2 \Big(4 \epsilon_{1} + \beta'^2 + \beta' \sqrt{4 \epsilon_{1} + \beta'^2}\Big) - k_{\parallel}^2 (4\epsilon_{1})} \hspace{0.2cm}\Bigg] \ ,
\end{equation}

and

\begin{equation}\label{eq:S-4}
k_{\perp(4)} = \frac{1}{\sqrt{4 \epsilon_{1} + \beta'^2}} \Bigg[- k_{\parallel} \beta' + \sqrt{\epsilon_{1} k_{0}^2 \Big(4 \epsilon_{1} + \beta'^2 + \beta' \sqrt{4 \epsilon_{1} + \beta'^2}\Big) - k_{\parallel}^2 (4\epsilon_{1})} \hspace{0.2cm}\Bigg] \ .
\end{equation}

Now setting $\beta'' = \frac{\beta'}{2}$ and $q^2 = \frac{k_{\parallel}^2}{k_{0}^2} = \sin^2{\theta_{in}} $, the four solutions given in Eq. (\ref{eq:S-1})--(\ref{eq:S-4}) can be further simplified to

\begin{equation} \label{eq:solution1}
\frac{k_{\perp(1)}}{k_{0}} = \frac{1}{2 \sqrt{1 + \beta''^2}} \Bigg[ q \beta' - 2 \sqrt{1 + \beta''^2 - \beta'' \sqrt{1 + \beta''^2} - q^2} \hspace{0.2cm}\Bigg] \ ,
\end{equation}

\begin{equation} \label{eq:solution2}
\frac{k_{\perp(2)}}{k_{0}} = \frac{1}{2 \sqrt{1 + \beta''^2}} \Bigg[ q \beta' + 2 \sqrt{1 + \beta''^2 - \beta'' \sqrt{1 + \beta''^2} - q^2} \hspace{0.2cm}\Bigg] \ ,
\end{equation}

\begin{equation} \label{eq:solution3}
\frac{k_{\perp(3)}}{k_{0}} = \frac{1}{2 \sqrt{1 + \beta''^2}} \Bigg[ -q \beta' - 2 \sqrt{1 + \beta''^2 + \beta'' \sqrt{1 + \beta''^2} - q^2} \hspace{0.2cm}\Bigg] \ ,
\end{equation}

and

\begin{equation} \label{eq:solution4}
\frac{k_{\perp(4)}}{k_{0}} = \frac{1}{2 \sqrt{1 + \beta''^2}} \Bigg[ -q \beta' + 2 \sqrt{1 + \beta''^2 + \beta'' \sqrt{1 + \beta''^2} - q^2} \hspace{0.2cm}\Bigg] \ .
\end{equation}

\begin{figure}[H]
\begin{center}
\includegraphics[scale=0.65]{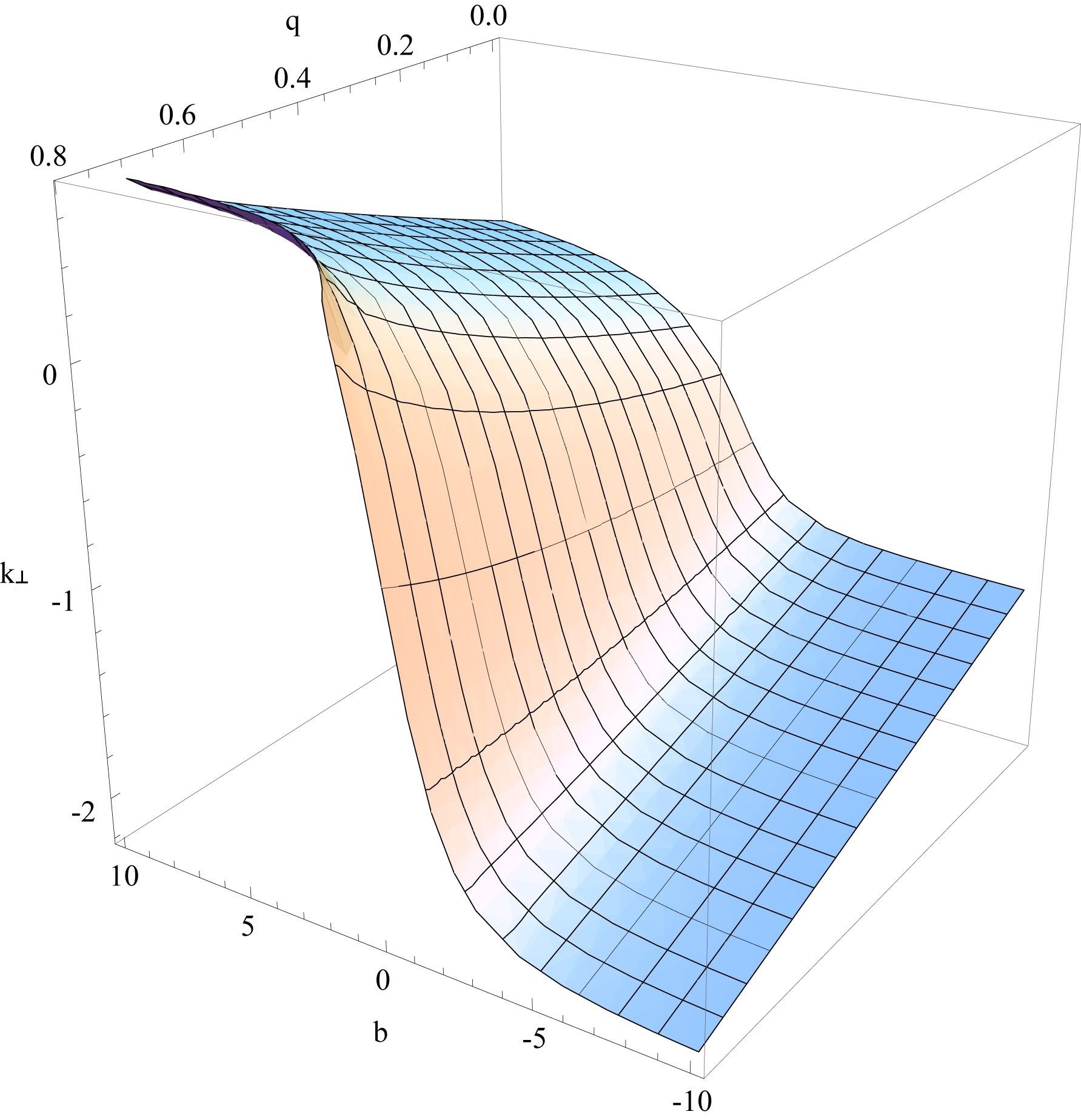}
\caption{Numerical plot of Eq. (\ref{eq:solution1}) with $\beta'' = \frac{\beta'}{2}= \frac{b}{2}$ and $q = \frac{k_{\parallel}}{k_{0}}$.}
\label{SW-sol-1}
\end{center}
\end{figure}

\begin{figure}[H]
\begin{center}
\includegraphics[scale=0.4]{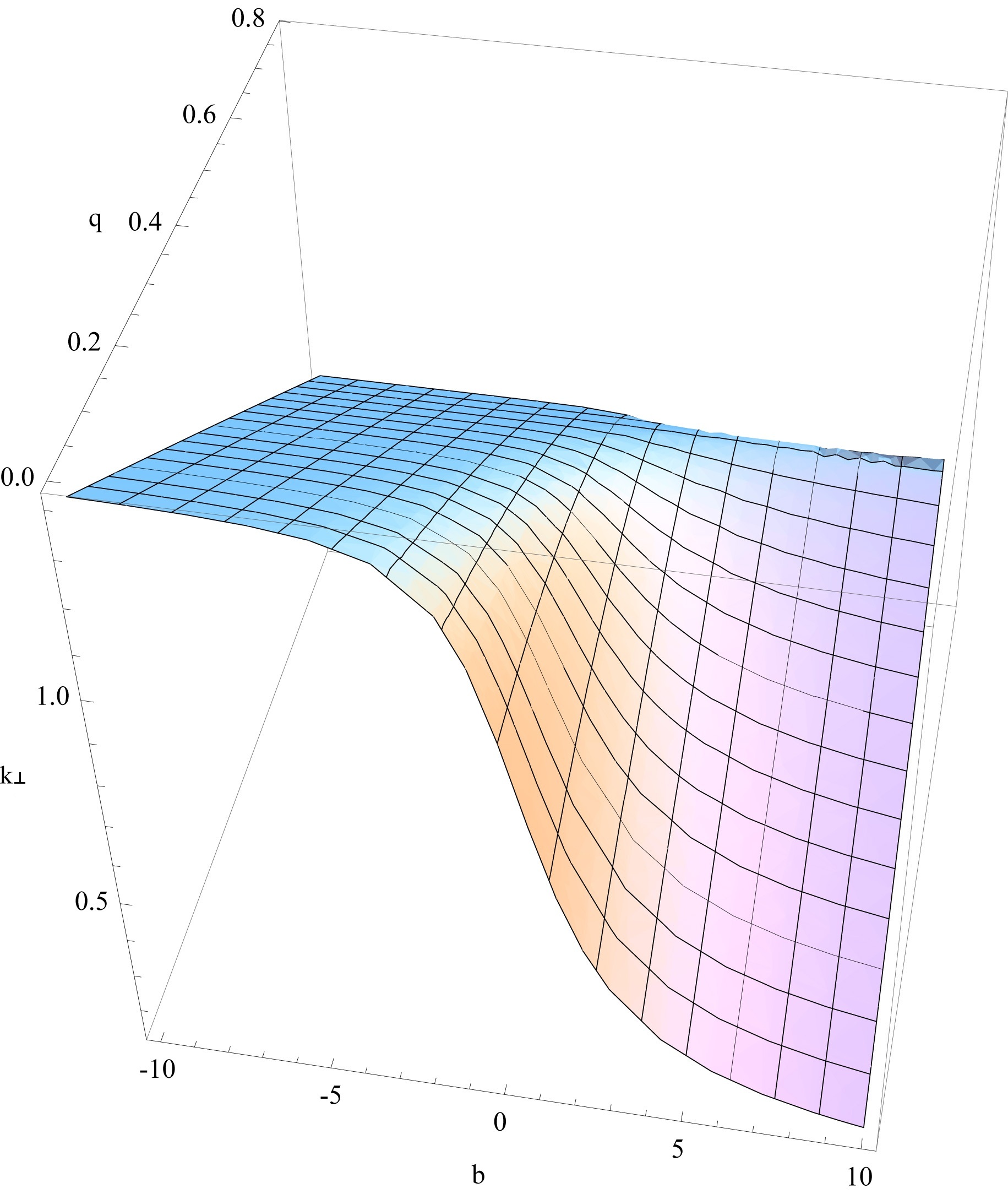}
\caption{Numerical plot of Eq. (\ref{eq:solution2}) with $\beta'' = \frac{\beta'}{2}= \frac{b}{2}$ and $q = \frac{k_{\parallel}}{k_{0}}$.}
\label{SW-sol-2}
\end{center}
\end{figure}

\begin{figure}[H]
\begin{center}
\includegraphics[scale=0.4]{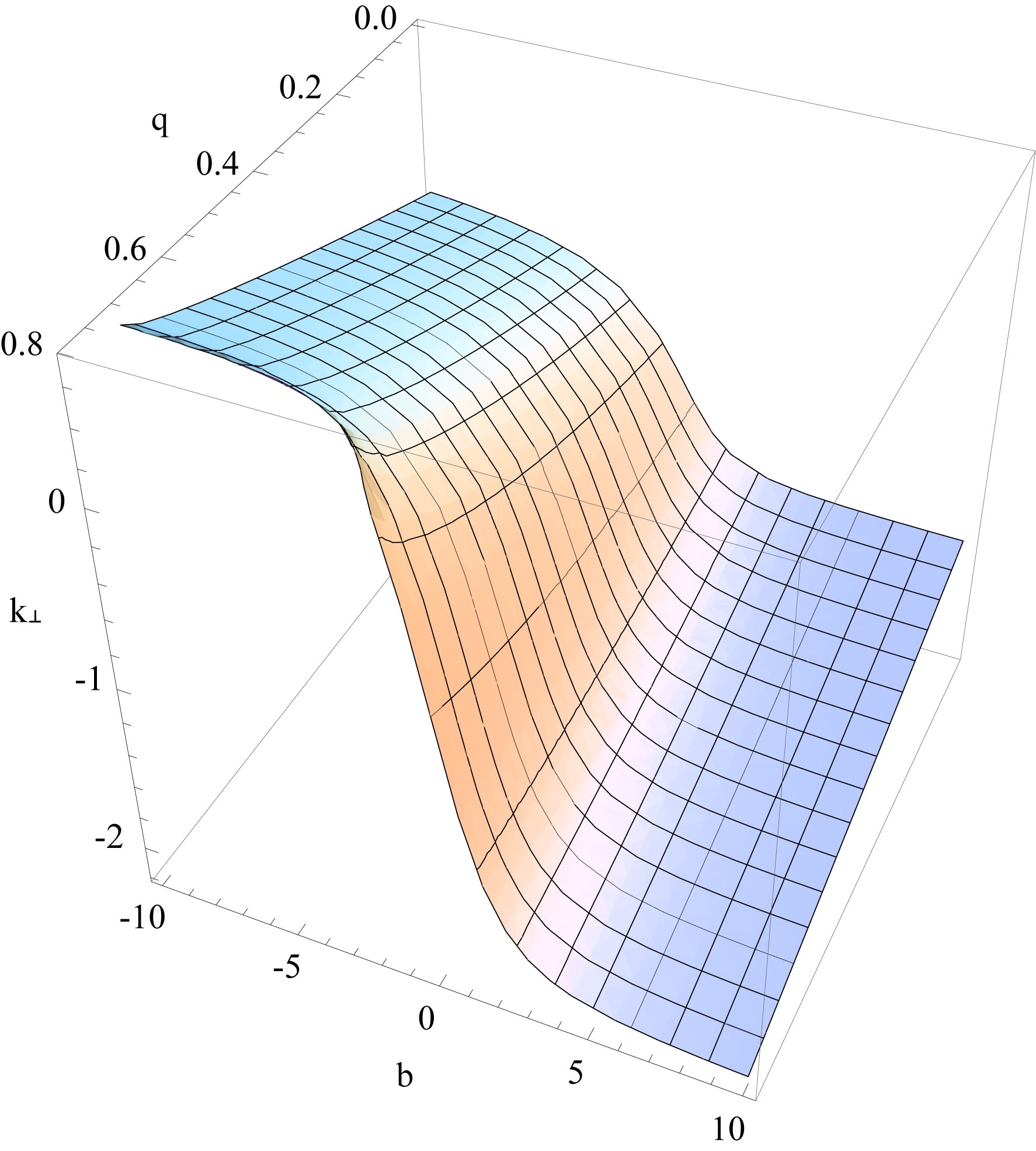}
\caption{Numerical plot of Eq. (\ref{eq:solution3}) with $\beta'' = \frac{\beta'}{2}= \frac{b}{2}$ and $q = \frac{k_{\parallel}}{k_{0}}$.}
\label{SW-sol-3}
\end{center}
\end{figure}

\begin{figure}[H]
\begin{center}
\includegraphics[scale=0.5]{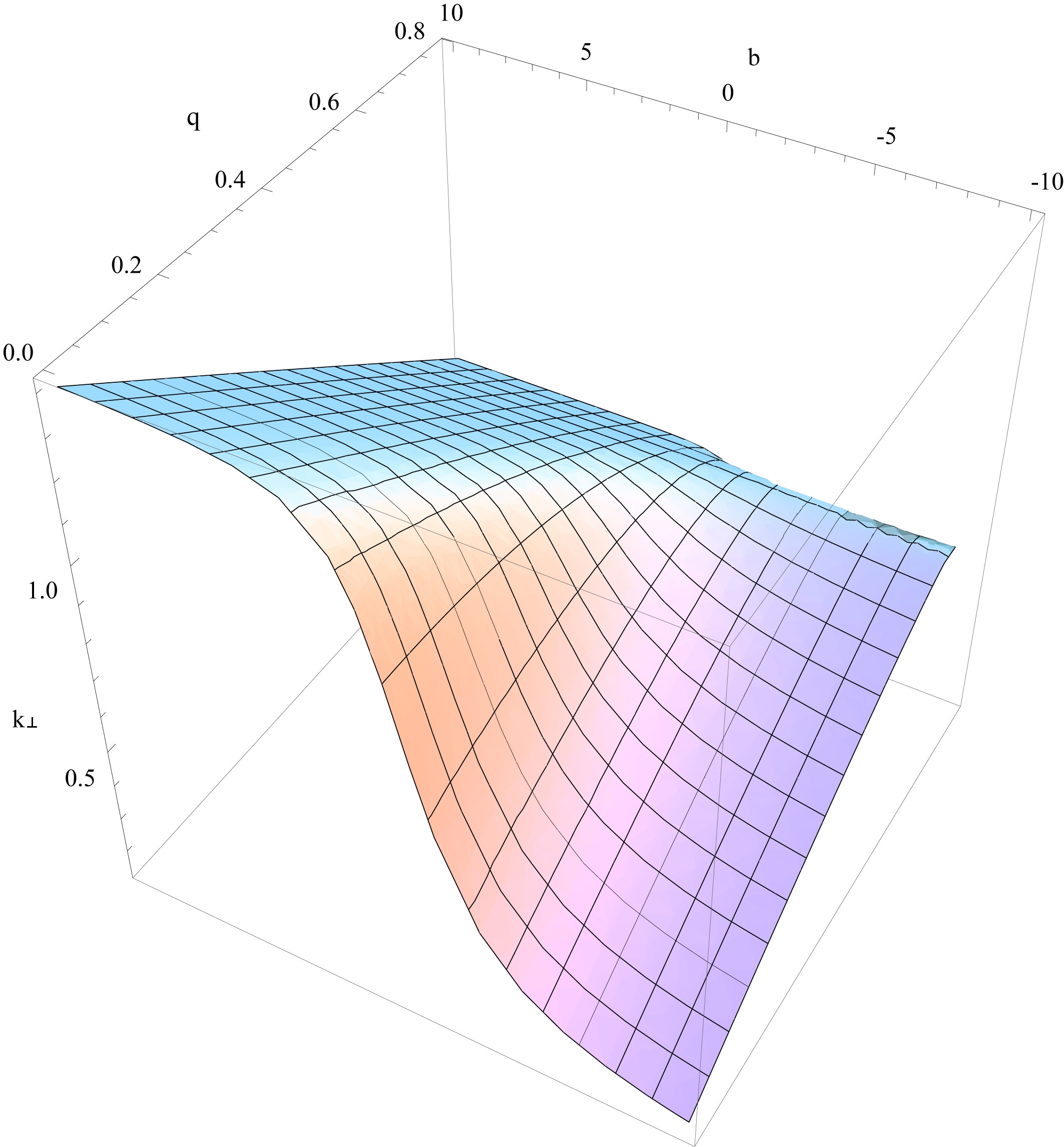}
\caption{Numerical plot of Eq. (\ref{eq:solution4}) with $\beta'' = \frac{\beta'}{2}= \frac{b}{2}$ and $q = \frac{k_{\parallel}}{k_{0}}$.}
\label{SW-sol-4}
\end{center}
\end{figure}

The Eq. (\ref{eq:solution1})--(\ref{eq:solution4}) show double Jones birefringence \cite{Roth}. From the solutions if the incident angle falls in the range

\begin{equation}
q^2 = \sin^2{\theta_{in}} > 1 + \beta'' \Big(\beta'' - \sqrt{(1+\beta''^2)}\Big) \ \ \ \textrm{for}  \ \ \beta'' > 0 \ ,
\end{equation}

it gives rise to complex $k_{\bot}$ that decays with \textit{z} to generate the surface wave. The visualization of the four solutions mentioned above is given in Figure. (\ref{SW-sol-1})--(\ref{SW-sol-4}). Figure. (\ref{SW-sol-1}) is the visualized surface for Eq. (\ref{eq:solution1}) generated in Mathematica. It is clear in the plot that the decay of the wave is constrained to $\beta'$. If we further increase the values of $\beta'$, the ends of the curve become more flat, however this does not influence the damping of the wave.

The crucial role of $\beta'$ as we have mentioned earlier turns out the same way as expected and for an incident electromagnetic wave on to the plane $z = 0$, it is playing a role in the damping of the wave. However, the surfaces generated using Mathematica for other solutions are slightly different. Figure. (\ref{SW-sol-2}) displays the surface generated for the Eq. (\ref{eq:solution2}). With an increase in the value of $q$, the $k_{\perp}$ component is decreasing and the depth of the curve is also becoming less. When the value of $q$ approaches nearly $0.8$, the surface is seen to be almost flat.

The solutions given in Eq. (\ref{eq:solution3}) and (\ref{eq:solution4}) are plotted in Figure. (\ref{SW-sol-3}) and (\ref{SW-sol-4}) respectively. A common behavior of damping for an electromagnetic wave is seen in both of the generated surface plots. These plots emphasize the role of $\beta'$ in our calculations for the uniaxial anisotropic magneto-electric medium and provide an illustration for the observation of physical phenomenon of surface waves in such a medium. This study on the magneto-electric medium using the additional axes in many aspects may give rise to the observation of more fundamental effects in optics.

\begin{figure}[H]
\begin{center}
\includegraphics[scale=0.5]{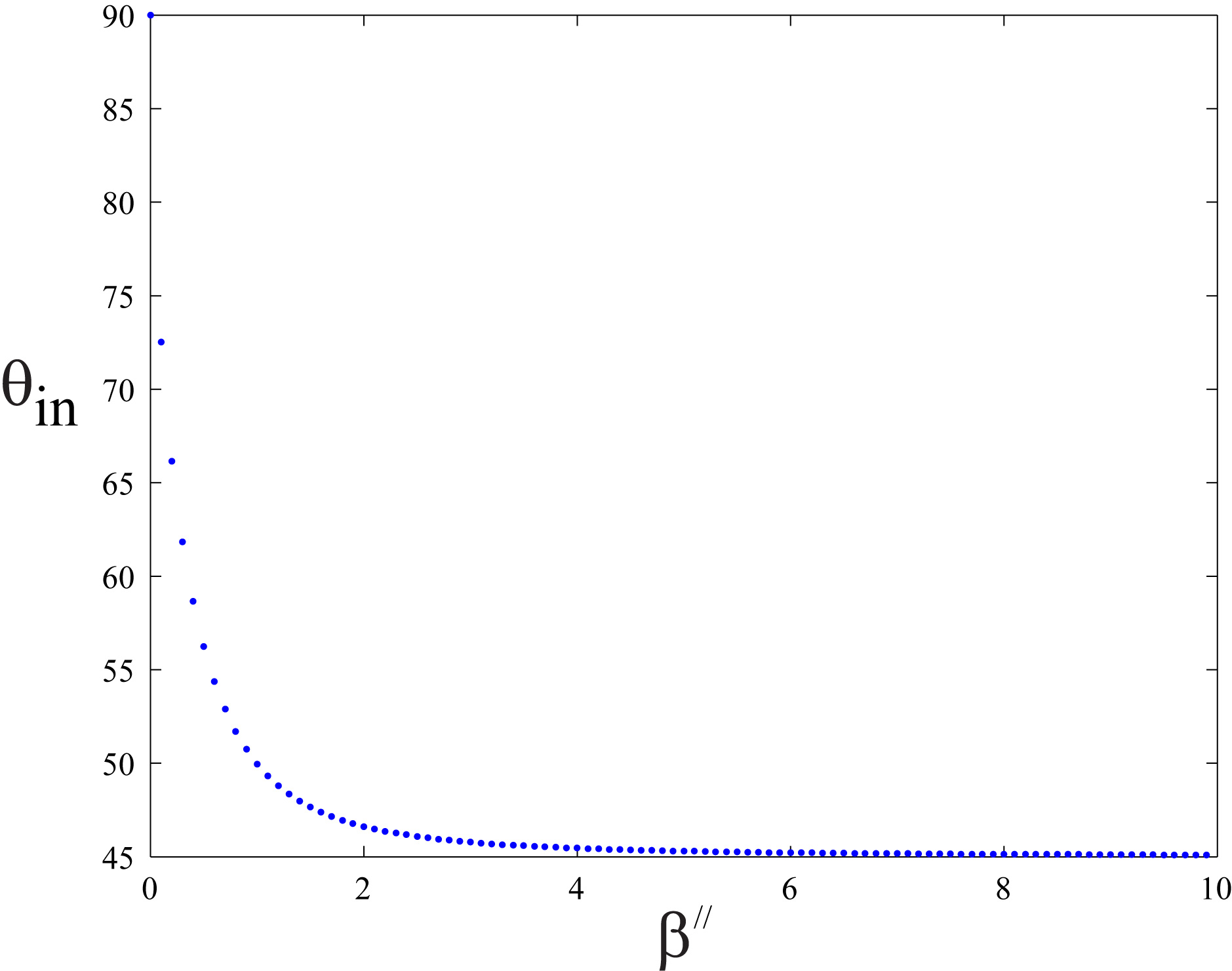}
\caption{The incident angle $\theta_{in}$ as a function of $\beta''$, where $\theta_{in}$ is in degrees.}
\label{theta-in}
\end{center}
\end{figure}

The behavior of the incident angle ($\theta_{in}$) with respect to $\beta''$ is shown in Figure. (\ref{theta-in}), where the angle is in degrees.

To make a comparison of the permittivity and permeability matrix with the tensorial form of ME based on the relativistic invarience \cite{FWHehl}, consider that $\epsilon$, $\alpha$ and $\mu$, all are diagonal with first two entries same. For such a case, using Eq. (\ref {eq:11}) the permittivity matrix simplifies to

\begin{equation}\label{epsilon-Hehl}
\tilde{\varepsilon}_{ij} = \Bigg( \begin{array}{ccc} \epsilon_{11}- \frac{\alpha_{11}^2}{\mu_{11}} & 0 & 0 \\
0 & \epsilon_{11}- \frac{\alpha_{11}^2}{\mu_{11}} & 0 \\
0 & 0 & \epsilon_{33}- \frac{\alpha_{33}^2}{\mu_{33}} \end{array} \Bigg) \ ,
\end{equation}

and the permeability matrix takes the form

\begin{equation}\label{mu-Hehl}
\mu_{ij}^{-1} = \Bigg( \begin{array}{ccc} \mu_{11}^{-1} & 0 & 0 \\
0 & \mu_{11}^{-1} & 0 \\
0 & 0 & \mu_{33}^{-1} \end{array} \Bigg) \ .
\end{equation}

Now consider another case when $\epsilon$, $\alpha$ and $\mu$, all are diagonal, however with different matrix elements. The Eq. (\ref{eq:10}) can be rewritten in the form

\begin{equation}
\alpha_{ik} = \beta_{im}\mu_{mk} \ .
\end{equation}

Using the definition of $\beta_{ij}$ from Eq. (\ref{eq:16}) into $\alpha_{ik}$, given in the last equation, we arrive at

\begin{equation}
\alpha_{ik} = \beta_{1} \delta_{im} \mu_{mk} + \beta' \hat d_{i} \hat d_{m} \mu_{mk} \ ,
\end{equation}

which for only $\hat d$, simplifies to

\begin{equation}\label{alpha-Hehl}
\Bigg( \begin{array}{ccc} \alpha_{11} & 0 & 0 \\
0 & \alpha_{22} & 0 \\
0 & 0 & \alpha_{33} \end{array} \Bigg) = \beta_{1} \Bigg( \begin{array}{ccc} \mu_{11}(1+\frac{\beta'}{\beta_{1}}\sin^2{\theta}\cos^2{\phi}) & 0 & 0 \\
0 & \mu_{22}(1+\frac{\beta'}{\beta_{1}} \sin^2{\theta} \sin^2{\phi}) & 0 \\
0 & 0 & \mu_{33}(1+ \frac{\beta'}{\beta_{1}}\cos^2{\theta}) \end{array} \Bigg) \ .
\end{equation}

In comparison with Ref. \cite{FWHehl} it depends on the orientation of the unit vector $\hat a$ in Eq. (\ref{eq:17}) that comes from $\tilde{\varepsilon}_{ij}$.

\section*{Conclusion}
\label{sec:C}
In this paper, we have solved the Maxwell's equations with the typical definition of the linear magneto-electric effect given in Eq. (\ref{MEeffectDi})--(\ref{MEeffectBi}), and extended the approach given in Ref. \cite{FVIgnatovich} to describe the propagation of an EM wave through a uniaxial anisotropic magnetic ME medium. By considering the linear ME effect, and by introducing the new basis, it is shown, that for a plain electromagnetic wave of the form $\vec E(\vec r) \ \textrm{exp} (- {\it{i}} \omega t)$, the polarization vector ($\vec{E}$) lies in the plane of two independent orthogonal vectors $\vec e_{1}$ and $\vec e_{2}$. Further proceeding with the calculations, it is observed, that the terms for the magneto-electric effect appear in the off diagonal place in the Eq. (\ref{eq:matrix1}), where the parameter $\beta'$ plays a crucial role. The solutions of the fourth order polynomial provide the condition for the physical phenomenon of D'yakonov surface wave at certain incident angle. The proposed surface wave is observable at an incident angle $\frac{\pi}{4}$. From the condition of the surface wave, it is inferred, that the constant $\beta'$ is responsible for the damping of surface wave as shown in the figures. Lastly, the Eq. (\ref{epsilon-Hehl}), (\ref{mu-Hehl}) and (\ref{alpha-Hehl}) support the Dzyaloshinskii's theory.\\


\section*{Acknowledgements}

The authors would like to thank National Science Foundation (NSF) of China ($11475088$ and $11275024$), and the Ministry of Science and Technology of China (Equipment-Project $2013$YQ$03059503$, and $863$Project $2013$AA$122901$) for their financial support for this work.

\section*{Author contributions}
W.M. and Q.Z. proposed the idea. W.M. performed the theoretical derivation and analysis. Q.Z. supervised the research. All authors contributed to the preparation of this manuscript.

\section*{Additional information}
The authors declare no competing financial interests.

\end{document}